\documentstyle[12pt,aaspp,epsf]{article}

\catcode`\@=11
\long\def\@makefntext#1{ %\parindent 1em
\protect\noindent \hbox to 3.2pt {\hskip-.9pt  
$^{{\ninerm\@thefnmark}}$\hfil}#1\hfill} %can be used 

\def\thefootnote{\fnsymbol{footnote}}
 \def\@makefnmark{\hbox to 0pt{$^{\@thefnmark}$\hss}}  %original 
	
\def\ps@myheadings{\let\@mkboth\@gobbletwo
\def\@oddhead{\hbox{} %\sl
\rightmark\hfil\ninerm\thepage}   
\def\@oddfoot{}\def\@evenhead{\ninerm\thepage\hfil %\sl
\leftmark\hbox{}}\def\@evenfoot{}
\def\sectionmark##1{}\def\subsectionmark##1{}}

\def\refitem #1! #2! #3! #4;{\hang\noindent
    \hangindent 20pt\rm #1, \rm #2, \rm #3, \rm #4.\par}
\def\bookref{\par\noindent\hangindent 20pt}

\def\wisk#1{\ifmmode{#1}\else{$#1$}\fi}
\def\Amp     {\wisk{{\langle Q_{RMS}^2\rangle^{0.5}}}}

\def\percm  {\wisk{{\rm cm}^{-1}}}

\def\lsim   {\wisk{_<\atop^{\sim}}}
\def\gsim   {\wisk{_>\atop^{\sim}}}

\def\Msun   {\wisk{{\rm M_\odot}}}

\def\COBE {{\it COBE}}
\def\SIRTF {{\it SIRTF}}

\def\etal {{\it et al.}}
\def\vs   {{\it vs.}}

\def\bc {\begin{center}}
\def\ec {\end{center}}

\def\be {\begin{equation}}
\def\ee {\end{equation}}

\begin{document}

%----------------------------PROCSLA.STY---------------------------------------
\newcommand{\symbolfootnote}{\renewcommand{\thefootnote}
	{\fnsymbol{footnote}}}
\renewcommand{\thefootnote}{\fnsymbol{footnote}}
\newcommand{\alphfootnote}
	{\setcounter{footnote}{0}
	 \renewcommand{\thefootnote}{\sevenrm\alph{footnote}}}

%------------------------------------------------------------------------------
%NEW DEFINED SECTION COMMANDS 
\newcounter{sectionc}\newcounter{subsectionc}\newcounter{subsubsectionc}
\renewcommand{\section}[1] {\vspace{0.6cm}\addtocounter{sectionc}{1} 
\setcounter{subsectionc}{0}\setcounter{subsubsectionc}{0}\noindent 
	{\bf\thesectionc. #1}\par\vspace{0.4cm}}
\renewcommand{\subsection}[1] {\vspace{0.6cm}\addtocounter{subsectionc}{1} 
	\setcounter{subsubsectionc}{0}\noindent 
	{\it\thesectionc.\thesubsectionc. #1}\par\vspace{0.4cm}}
\newcommand{\nonumsection}[1] {\vspace{0.6cm}\noindent{\bf #1}
	\par\vspace{0.4cm}}
					         
%NEW MACRO TO HANDLE APPENDICES
\newcounter{appendixc}
\newcounter{subappendixc}[appendixc]
\newcounter{subsubappendixc}[subappendixc]
\renewcommand{\thesubappendixc}{\Alph{appendixc}.\arabic{subappendixc}}
\renewcommand{\thesubsubappendixc}
	{\Alph{appendixc}.\arabic{subappendixc}.\arabic{subsubappendixc}}

\renewcommand{\appendix}[1] {\vspace{0.6cm}
        \refstepcounter{appendixc}
        \setcounter{figure}{0}
        \setcounter{table}{0}
        \setcounter{equation}{0}
        \renewcommand{\thefigure}{\Alph{appendixc}.\arabic{figure}}
        \renewcommand{\thetable}{\Alph{appendixc}.\arabic{table}}
        \renewcommand{\theappendixc}{\Alph{appendixc}}
        \renewcommand{\theequation}{\Alph{appendixc}.\arabic{equation}}
%       \noindent{\bf Appendix \theappendixc. #1}\par\vspace{0.4cm}}
        \noindent{\bf Appendix \theappendixc #1}\par\vspace{0.4cm}}
\newcommand{\subappendix}[1] {\vspace{0.6cm}
        \refstepcounter{subappendixc}
        \noindent{\bf Appendix \thesubappendixc. #1}\par\vspace{0.4cm}}
\newcommand{\subsubappendix}[1] {\vspace{0.6cm}
        \refstepcounter{subsubappendixc}
        \noindent{\it Appendix \thesubsubappendixc. #1}
	\par\vspace{0.4cm}}

%------------------------------------------------------------------------------
%MACRO FOR ABSTRACT BLOCK
\def\abstracts#1{{
	\centering{\begin{minipage}{30pc}\tenrm\baselineskip=12pt\noindent
	\centerline{\tenrm ABSTRACT}\vspace{0.3cm}
	\parindent=0pt #1
	\end{minipage} }\par}} 

%------------------------------------------------------------------------------
%NEW MACRO FOR BIBLIOGRAPHY
\newcommand{\bibit}{\it}
\newcommand{\bibbf}{\bf}
\renewenvironment{thebibliography}[1]
	{\begin{list}{\arabic{enumi}.}
	{\usecounter{enumi}\setlength{\parsep}{0pt}
%1.25cm IS STRICTLY FOR PROCSLA.TEX ONLY
\setlength{\leftmargin 1.25cm}{\rightmargin 0pt}
%0.52cm IS FOR NEW DATA FILES
%\setlength{\leftmargin 0.52cm}{\rightmargin 0pt}
	 \setlength{\itemsep}{0pt} \settowidth
	{\labelwidth}{#1.}\sloppy}}{\end{list}}

%------------------------------------------------------------------------------
%FOLLOWING THREE COMMANDS ARE FOR 'LIST' COMMAND.
\topsep=0in\parsep=0in\itemsep=0in
\parindent=1.5pc

%LIST ENVIRONMENTS
\newcounter{itemlistc}
\newcounter{romanlistc}
\newcounter{alphlistc}
\newcounter{arabiclistc}
\newenvironment{itemlist}
    	{\setcounter{itemlistc}{0}
	 \begin{list}{$\bullet$}
	{\usecounter{itemlistc}
	 \setlength{\parsep}{0pt}
	 \setlength{\itemsep}{0pt}}}{\end{list}}

\newenvironment{romanlist}
	{\setcounter{romanlistc}{0}
	 \begin{list}{$($\roman{romanlistc}$)$}
	{\usecounter{romanlistc}
	 \setlength{\parsep}{0pt}
	 \setlength{\itemsep}{0pt}}}{\end{list}}

\newenvironment{alphlist}
	{\setcounter{alphlistc}{0}
	 \begin{list}{$($\alph{alphlistc}$)$}
	{\usecounter{alphlistc}
	 \setlength{\parsep}{0pt}
	 \setlength{\itemsep}{0pt}}}{\end{list}}

\newenvironment{arabiclist}
	{\setcounter{arabiclistc}{0}
	 \begin{list}{\arabic{arabiclistc}}
	{\usecounter{arabiclistc}
	 \setlength{\parsep}{0pt}
	 \setlength{\itemsep}{0pt}}}{\end{list}}

%------------------------------------------------------------------------------
%FIGURE CAPTION
\newcommand{\fcaption}[1]{
        \refstepcounter{figure}
        \setbox\@tempboxa = \hbox{\tenrm Fig.~\thefigure. #1}
        \ifdim \wd\@tempboxa > 6in
           {\begin{center}
        \parbox{6in}{\tenrm\baselineskip=12pt Fig.~\thefigure. #1 }
            \end{center}}
        \else
             {\begin{center}
             {\tenrm Fig.~\thefigure. #1}
              \end{center}}
        \fi}

%TABLE CAPTION
\newcommand{\tcaption}[1]{
        \refstepcounter{table}
        \setbox\@tempboxa = \hbox{\tenrm Table~\thetable. #1}
        \ifdim \wd\@tempboxa > 6in
           {\begin{center}
        \parbox{6in}{\tenrm\baselineskip=12pt Table~\thetable. #1 }
            \end{center}}
        \else
             {\begin{center}
             {\tenrm Table~\thetable. #1}
              \end{center}}
        \fi}

%------------------------------------------------------------------------------
%ACKNOWLEDGEMENT: this portion is from John Hershberger
\def\@citex[#1]#2{\if@filesw\immediate\write\@auxout
	{\string\citation{#2}}\fi
\def\@citea{}\@cite{\@for\@citeb:=#2\do
	{\@citea\def\@citea{,}\@ifundefined
	{b@\@citeb}{{\bf ?}\@warning
	{Citation `\@citeb' on page \thepage \space undefined}}
	{\csname b@\@citeb\endcsname}}}{#1}}

\newif\if@cghi
\def\cite{\@cghitrue\@ifnextchar [{\@tempswatrue
	\@citex}{\@tempswafalse\@citex[]}}
\def\citelow{\@cghifalse\@ifnextchar [{\@tempswatrue
	\@citex}{\@tempswafalse\@citex[]}}
\def\@cite#1#2{{$\null^{#1}$\if@tempswa\typeout
	{IJCGA warning: optional citation argument 
	ignored: `#2'} \fi}}
\newcommand{\citeup}{\cite}

%------------------------------------------------------------------------------
%FOR FNSYMBOL FOOTNOTE AND ALPH{FOOTNOTE} 
\def\fnm#1{$^{\mbox{\scriptsize #1}}$}
\def\fnt#1#2{\footnotetext{\kern-.3em
	{$^{\mbox{\sevenrm #1}}$}{#2}}}

%------------------------------------------------------------------------------
\font\twelvebf=cmbx10 scaled\magstep 1
\font\twelverm=cmr10 scaled\magstep 1
\font\twelveit=cmti10 scaled\magstep 1
\font\elevenbfit=cmbxti10 scaled\magstephalf
\font\elevenbf=cmbx10 scaled\magstephalf
\font\elevenrm=cmr10 scaled\magstephalf
\font\elevenit=cmti10 scaled\magstephalf
\font\bfit=cmbxti10
\font\tenbf=cmbx10
\font\tenrm=cmr10
\font\tenit=cmti10
\font\ninebf=cmbx9
\font\ninerm=cmr9
\font\nineit=cmti9
\font\eightbf=cmbx8
\font\eightrm=cmr8
\font\eightit=cmti8
%----------------------START OF DATA FILE------------------------------

\noindent
\makebox[0pt][l]{
\raisebox{36pt}[0pt][0pt]{astro-ph/9408002, UCLA-ASTRO-ELW-94-01}}

\centerline{\tenbf DARK MATTER IN THE LIGHT OF COBE\,\footnotemark[1]
\footnotetext[1]{The National Aeronautics and Space Administration/Goddard
Space Flight Center (NASA/GSFC) is responsible for the design, development,
and operation of the Cosmic Background Explorer (\COBE).
Scientific guidance is provided by the \COBE\ Science Working Group.
GSFC is also responsible for the development of the analysis software
and for the production of the mission data sets.}
}

\pagestyle{empty}

\vspace{0.8cm}

\centerline{\tenrm EDWARD L. WRIGHT}
\baselineskip=13pt
\centerline{\tenit UCLA Astronomy}
\baselineskip=12pt
\centerline{\tenit Los Angeles CA 90024-1562}

\abstracts{The observations of all three \COBE\
instruments are examined for the effects of dark matter.
The anisotropy measured by the DMR, and especially the degree-scale
ground- and balloon-based experiments, is only compatible with large-scale
structure formation by gravity if the Universe is dominated by non-baryonic
dark matter.
The FIRAS instrument measures the total power radiated by cold dust, and thus 
places tight limits on the absorption of starlight by very cold gas and dust
in the outer Milky Way.
The DIRBE instrument measures the infrared background, and will place tight
limits on the emission by low mass stars in the Galactic halo.
}

\parindent 20pt
\baselineskip 14pt

\section{Introduction}

While \COBE\ (Boggess \etal\ 1992)
has no instruments that directly detect dark matter,
its three instruments offer important clues about the baryonic and 
non-baryonic content of the Universe.  The FIRAS observations of the
spectrum of the cosmic microwave background radiation (CMBR) show
that any deviation from a blackbody are very small
(Mather \etal\ 1990 and Mather \etal\ 1994).  This limits the possible
effect of energetic explosions on the formation of large-scale structure
(Wright \etal\ 1994).  If gravity is the force responsible for
large-scale structure, then the DMR observations of anisotropy
require a non-baryonic dark matter dominated Universe.  Even the baryons
in the Universe are mostly in a dark form, but FIRAS observations of the
millimeter emission from the Galaxy show that these dark baryons can not be
in clouds of very cold gas and dust associated with the CO absorbing clouds
seen by Lequeux \etal\ (1993).  However, even more compact configurations of
baryons are allowed: brown dwarfs.  A Galactic halo of old cold brown dwarfs
will be essentially undetectable by the DIRBE instrument unless all of the
mass is in objects right at the limit of hydrogen burning.

\newpage

\section {DMR $\Delta T$ and non-Baryonic Dark Matter}

\begin{figure}[t]
\plotone{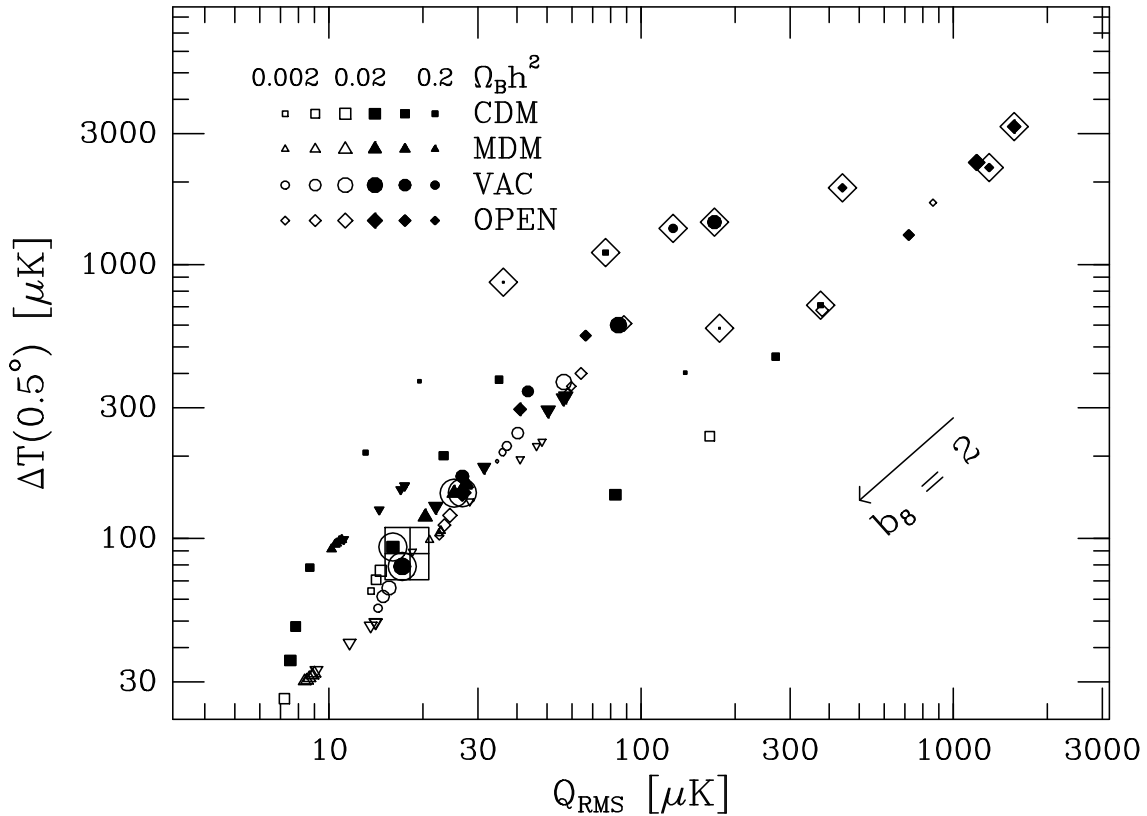}
\caption{Predicted $\Delta T$ for Holtzman models at 0.5$^\circ$ scale
\vs\ quadrupole scale.}
\label{baryon}
\end{figure}

The DMR anisotropy implies a small level of gravitational potential
perturbations via the Sachs-Wolfe (1967) effect.  At the same time, models of
large-scale structure formation require certain levels of gravitational forces
which can be converted into predicted $\Delta T$'s.  Figure \ref{baryon}
shows the predictions from the models of Holtzman (1989) compared to the \COBE\
DMR \Amp\ and the anisotropy at 0.5$^\circ$ measured by the MAX
experiment (Clapp \etal\ 1994 and Devlin \etal\ 1994).  The models with only
baryonic matter are surrounded by open diamonds, while models emphasized by
Wright \etal\ (1992) are surrounded by open circles.  The CDM+baryon model and
the vacuum dominated model (which still has 90\% of the matter non-baryonic)
both sit on top of the observed $\Delta T$'s, while the open model and the
mixed dark matter model need bias factors $b_8 < 2$ to agree with the data.
The nearest baryonic model needs $b_8 \approx 10$ to fit the data, which is not
reasonable.  This problem with baryonic models arises because non-baryonic dark
matter perturbations start to grow at $z_{eq} \approx 6000$, while baryonic
perturbations can only start to grow at $z_{rec} \approx 10^3$, and thus lose a
factor of $\gsim 6$ in growth.

\section{FIRAS Limits on Very Cold Dust}

\begin{figure}[t]
\plotone{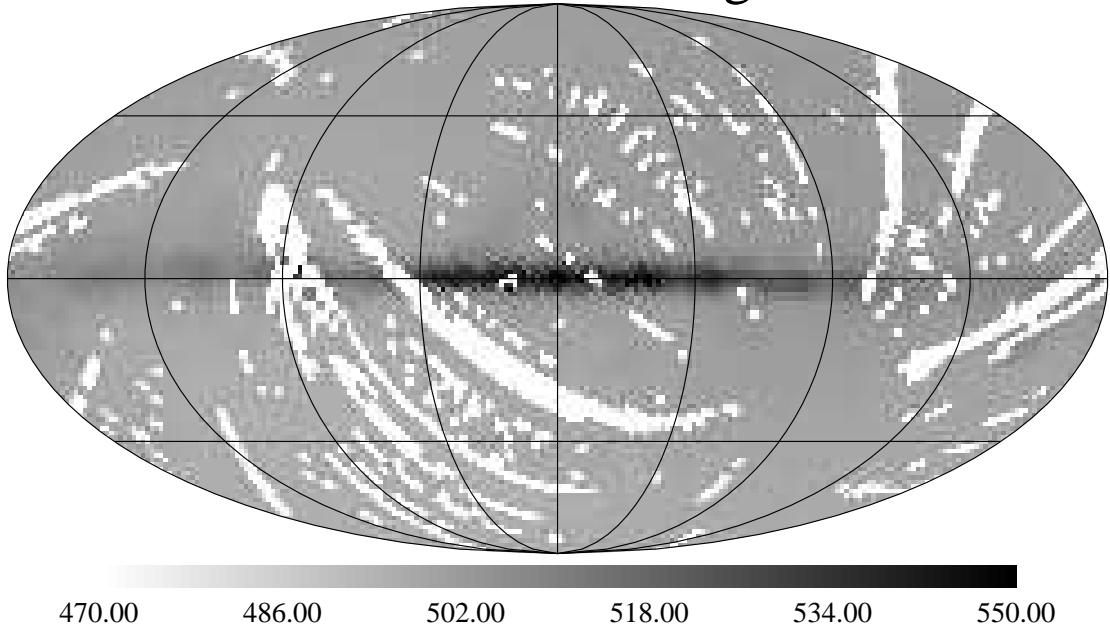}
\caption{The total FIRAS flux in the 8-14/cm band.}
\label{total814}
\end{figure}

\begin{figure}[t]
\plotone{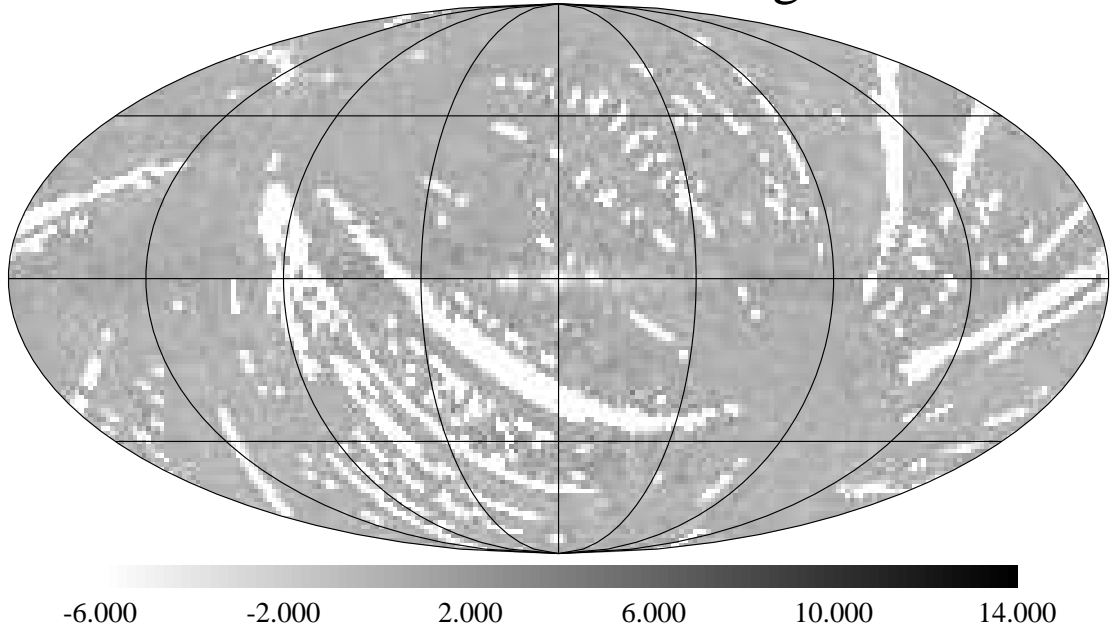}
\caption{The residual FIRAS flux after subtracting the monopole and dipole and
fitting for Galactic dust.}
\label{resid814}
\end{figure}

Lequeux \etal\ (1993) have observed CO absorption toward extragalactic radio
sources at low galactic latitudes.  In at least one case the CO emission is
very low, indicating very cold gas with $T_{ex} = 3.5$~K.  Since most surveys
for interstellar material are based on measuring emission, there is the
possibility that a large amount of very cold gas and dust could be hidden in
cold clouds.  But any starlight absorbed by these clouds will be reradiated in
the millimeter region where the FIRAS instrument on \COBE\ is sensitive.  FIRAS
uses bolometers which are approximately equally sensitive at all wavelengths,
so the limit placed on the absorption of starlight by the cold clouds is
roughly independent of their temperatures.  Because of the measured
$T_{ex} = 3.5$~K, I have chosen the band $8-14\;\percm$ which corresponds
to $h\nu/kT_{ex} = 3.3$ to 5.8, which should include a fraction 
\be
\frac{\int_8^{14}
\nu^\beta\left[B_\nu(T_d)-B_\nu(T_\circ)\right] d\nu}
{\int_0^\infty \nu^\beta\left[B_\nu(T_d)-B_\nu(T_\circ)\right] d\nu} 
\approx 0.5
\ee
of the power power emitted by the very cold dust grains with
emissivities varying like $\nu^\beta$ for $\beta \approx 1.5$
and dust temperature $T_d = 3.5$~K.

Lequeux \etal\ observe 4 clouds on a total path length $\sum \csc |b| = 66$.
Thus the optical depth from pole to pole of the disk is 
$\tau = 2\times 4/66 = 0.12$.
This implies that a plane-parallel galactic disk in an isotropic
optical background $J$ will absorb a power per unit area of 
\begin{eqnarray}
P & = & J \int \sin|b| (1-\exp(-\tau\csc|b|)) \cos b db dl \nonumber\\
  & = & 4\pi J \int_0^1 \mu (1-\exp(-\tau/\mu)) d\mu = 1.24 J.
\end{eqnarray}
Considering only the part of the sky with $\sin |b| < 0.1$, I expect that
the flux per radian of Galactic plane observed by FIRAS will be
approximately
\be
F = \frac{P\ln(s_{max}/s_{min})}{4\pi}
\ee
where the maximum distance along the line-of-sight $s_{max}$
can be taken as the
exponential scale length of the disk, or about 3 kpc, while the minimum
distance $s_{min}$
can be taken to be 10 times the scale height of the disk or about 1
kpc.  Thus the very cold dust should radiate about
$0.05 J$ W/m$^2$/rad into the 8-14 \percm\ band.

The mean excess emission in the 8-14 \percm\ band over the model
\be
I_\nu(l,b) = 
B_\nu(T_\circ + D_x \cos l \cos b + D_y \sin l \cos b + D_z \sin b)
+ g(\nu) G(l,b)
\ee
in the region with $\sin |b| < 0.1$ 
and $\cos l < 0.866$ is $6 \times 10^{-10}$ W/m$^2$/sr,
where $T_\circ$ is the mean temperature of the
cosmic background, $D_i$ are the components of the dipole
anisotropy, $g(\nu)$ is the average galactic dust 
spectrum, and $G(l,b)$ is the dust map (Wright \etal\ 1991).
This model fixes the monopole and dipole using the high galactic latitude
values, but allows for an adjustable amount of low latitude ``normal''
dust based on the 2-20 \percm\ spectrum (rather than using DIRBE
or FIRAS data at higher frequencies),
so there is only one parameter at each pixel.
The fit removes some of the 8-14 \percm\ power but for dust with
$T_d = 3.5$~K and power law emissivities $\propto \nu^\beta$ with 
$\beta$ between 0 and 1.8, the fraction of the total very cold dust emission
within the 8-14 \percm\ band after the fit is 
\be
\frac{\int_8^{14}
\left\{\nu^\beta\left[B_\nu(T_d)-B_\nu(T_\circ)\right] - G g(\nu) 
\right\}d\nu}
{\int_0^\infty \nu^\beta\left[B_\nu(T_d)-B_\nu(T_\circ)\right] d\nu} 
\approx 0.3
\ee
with
\be
G = \frac
{\sum_i 
\nu_i^\beta\left[B_{\nu_i}(T_d)-B_{\nu_i}(T_\circ)\right] 
g(\nu_i)/\sigma_i^2}
{\sum_i g(\nu_i)^2/\sigma_i^2}
\ee
using $\nu_i$, $g(\nu_i)$ and $\sigma_i$ from Mather \etal\ (1994).
This reduces the expected excess to $0.03 J$ W/m$^2$/rad.

Converting the observed excess 
to a flux per radian requires multiplying by the range of latitude used,
which is $\Delta b = 0.2$~rad.  The result is
\be
0.03 J = (0.2 \; \hbox{\rm rad}) \times (6 \times 10^{-10} \; 
\hbox{\rm W/m$^2$/sr}) = 1.2 \times 10^{-11} \hbox{\rm W/m$^2$/rad}
\ee 
so I obtain $J = 4 \times 10^{-9}$~W/m$^2$/sr.  The total power in the
2.7 K background is $10^{-6}$~W/m$^2$/sr so this limit is 0.4\% of the
CMB energy.  The interstellar radiation field (ISRF) given by Wright (1993)
integrates to $2.5 \times 10^{-6}$~W/m$^2$/sr, which is 600 times higher
than $J$.  Thus if I assume that the CO clouds seen by Lequeux \etal\ are
optically thick to starlight, and that the
local ISRF decays outward with the normal exponential disk scale length
of 3 kpc, then the predicted millimeter excess from the
Galactic plane in the 8-14 \percm\ range should be 600 times higher than it
is.  Even the heating provided by the diffuse extragalactic light (DEGL)
would produce an 8-14 \percm\ excess that is 10 times higher than the 
observed excess.

I conclude that CO clouds observed by Lequeux \etal\ can not be extremely
optically thick objects hiding a large mass of very cold gas and dust.
The observed number of clouds per unit $\csc b$ implies an absorption of
the ISRF and DEGL that would produce a very large signal in the FIRAS
Galactic plane observations, while only a small signal is seen.  The low
excitation temperature in CO is probably due to low density, giving
$T_\circ \lsim T_{ex} << T_{gas} \approx T_{dust}$.  
The clouds do absorb starlight, but the
dust has a normal dust temperature, and the reradiation is already included in
the FIRAS and DIRBE observations of the Galaxy.

\newpage

\section{DIRBE Limits on a Brown Dwarf Halo}

\begin{figure}[t]
\plotone{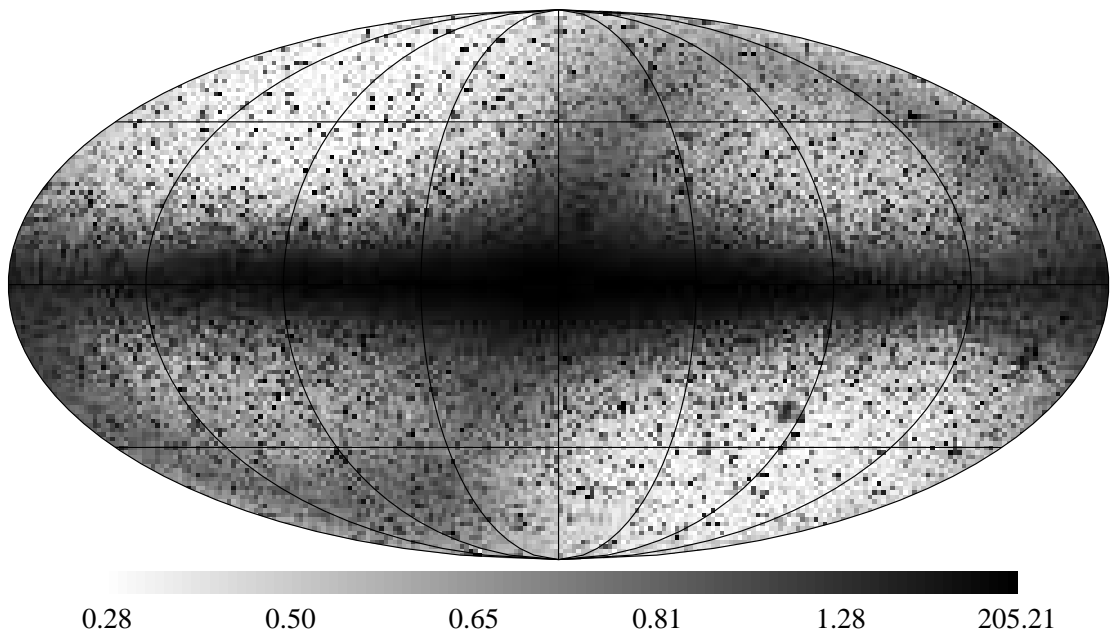}
\caption{DIRBE flux in arbitrary linear units at 3.5 $\mu$m. 
The darkest part of the sky is $\approx 10^5$ Jy/sr
(Hauser 1994).}
\label{dirbe3}
\end{figure}

Adams \& Walker (1990) and Daly \& McLaughlin (1992)
have computed the expected intensity from a brown
dwarf Galactic halo.  The expected flux is generally very low unless all
of the halo density is made up of maximum mass brown dwarfs,
$M = 0.08 \Msun$.
The halo mass density assumed by Adams \& Walker is
\be
\rho(r) = \rho_\circ \frac{a^2}{a^2+r^2}
\ee
with $a = 2$~kpc and $\rho_\circ = 0.19\;$M$_\odot$/pc$^3$ 
which gives a local density of $0.01\;$M$_\odot$/pc$^3$.  
The mass column density to the galactic
pole is 137 M$_\odot$/pc$^2$ in this model.  
To convert this into a flux I need
the mass and luminosity of a brown dwarf.  I will assume $M = 0.05\;$M$_\odot$,
and the Burrows, Hubbard \& Lunine (1989) luminosity of $10^{-6}\;$L$_\odot$
with an effective temperature of 632 K at an age of 10 Gyr.  With an $M/L$ of
50,000 the resulting intensity at the pole is
$137 \;$M$_\odot$/pc$^2/(4\pi M/L) = 2.2 \times 10^{-4} \;$L$_\odot$/pc$^2$/sr 
$ = 10^{-10}$~W/m$^2$/sr.  
If I assume that the spectrum is a blackbody then the flux in
individual bands is easy to find.  The ratio $\nu F_\nu /F_{bol} =
(15/\pi^4) x^4/(e^x-1)$ with $x = h\nu/kT$, which is 0.4 for
$x = 6.5$ at 3.5 $\mu$m with T = 632 K.  Thus the intensity is
48 Jy/sr in this model.  Since the total brightness at the South Ecliptic Pole
is 175 kJy/sr, the prospects for detecting an 0.03\% effect due to brown dwarfs
are not very good.

Note, however, that the assumption of a blackbody spectrum is likely to be very
bad.  The spectrum of Jupiter shows very large features in the infrared. 
Warmer brown dwarfs with dusty atmospheres will have smoother spectra
if the dust has a power law opacity $\kappa_\nu \propto \nu^\beta$.
Since dust absorbs more at short wavelengths than at long wavelengths, the
color temperature of a dusty brown dwarf will be smaller than the effective
temperature, leading to an apparent emissivity 
$\epsilon = (T_{eff}/T_{color})^4$ that is close to 2.  
Since the emissivity is used to estimate the size of brown dwarfs, calculations
based on blackbody emission will lead to sizes over-estimated by a factor of
about 1.4.
The lowered color temperature will further
reduce the expected 3.5 $\mu$m intensity, 
and it also makes old cold brown dwarfs
almost impossible to find in 2.2 $\mu$m surveys like those of 
Cowie \etal\ (1990).  
However, the individual brown dwarfs can easily be detected by the proposed
Space InfraRed Telescope Facility (\SIRTF).
In a \SIRTF\ $5^\prime \times 5^\prime$ field of view,
the closest brown dwarf will be 66 pc away, and produce a flux of
3.7 $\mu$Jy at 3.5 $\mu$m, but only 0.32 $\mu$Jy at 2.2 $\mu$m.
The lowered color temperature expected for dusty brown dwarfs
will give fluxes of
2.2 $\mu$Jy at 3.5 $\mu$m, but only 0.09 $\mu$Jy at 2.2 $\mu$m.
\SIRTF\ will have a sensitivity of 0.02 $\mu$Jy per pixel at 3.5 $\mu$m in
a 2500 second observation and will easily detect many brown dwarfs per FOV
at 3.5 $\mu$m, even though the integrated intensity is too small to be
seen by DIRBE.

\section{Summary}

The observations by \COBE\ of the CMBR show no evidence for non-gravitational
forces producing large-scale structure.  The gravitational forces implies by
the DMR measurements of $\Delta T$ are sufficient to produce the observed
large-scale structure only if most of the matter in the Universe responds
freely to these gravitational forces before recombination, which requires
non-baryonic dark matter.  The baryonic dark matter cannot be very cold gas
and dust associated with the CO absorption lines seen by Lequeux because
it would produce too much millimeter wave emission from the Galactic plane.
However, the dark baryons can easily be brown dwarfs which will escape 
detection by \COBE\ and ground-based IR surveys
but may well be seen by \SIRTF.

\section{References}

\refitem
Adams, F. C. \& Walker, T. P. 1990! ApJ! 359! 57-62;

\refitem
Boggess, N.~W. \etal\ 1992! ApJ! 397! 420-429;
%.!``The COBE Mission: Its Design and Performance Two Years After Launch'',

\refitem
Burrows, A., Hubbard, W. B \& Lunine, J. I. 1989! ApJ! 345! 939-958;
%.!"Theoretical Models of Very Low Mass Stars and Brown Dwarfs"

\refitem
Clapp, A. C., Devlin, M. J., Gundersen, J. O., Hagmann, C. A.,
Hristov, V. V., Lange, A. E., Lim, M., Lubin, P. M., Mauskopf, P. D.,
Meinhold, P. R., Richards, P. L., Smoot, G. F., Tanaka, S. T.,
Timbie, P. T. \& Wuensche, C. A. 1994! ApJL! TBD! TBD;
%.!"Measurements of Anisotropy in the Cosmic Microwave Background Radiation at
%.! Degree Angular Scales Near the Stars Sigma Hercules and Iota Draconis"

\refitem
Cowie, L. L., Gardner, J. P., Lilly, S. J. \& McLean, I. 1990!
ApJL! 360! L1-L4;
%.!"A K Band Deep Survey"

\refitem
Daly, R. A. \& McLaughlin, G. C. 1992! ApJ! 390! 423-430;
%.!"Dark matter and Brown Dwarfs: Prospect for the Direct Detection of a
%.! Brown Dwarf Halo".

\refitem
Devlin, M. J., Clapp, A. C., Gundersen, J. O., Hagmann, C. A.,
Hristov, V. V., Lange, A. E., Lim, M., Lubin, P. M., Mauskopf, P. D.,
Meinhold, P. R., Richards, P. L., Smoot, G. F., Tanaka, S. T.,
Timbie, P. T. \& Wuensche, C. A. 1994! ApJL! 430! L1-L4;
%.!"Measurements of Anisotropy in the Cosmic Microwave Background Radiation
%.! at 0.5 Degree Angular Scales Near the Star Gamma Ursae Minoris"

\bookref
Hauser, M. G. 1994, presented at the Extragalactic Background
Radiation Symposium at STScI, COBE Preprint 94-13.

\refitem
Holtzman, J. 1989! ApJSupp! 71! 1-24;
%.!"Microwave Background Anisotropies and Large Scale Structure in Universes
%.! with CDM, Baryons, Radiation and Massive and Massless Neutrinos"

\refitem
Lequeux, J., Allen, R. J. \& Guilloteau, S. 1993! A\&A! 280! L23-L26;
%.!"CO absorption in the outer Galaxy: abundant cold molecular gas"

\refitem
Mather, J.~C., Cheng, E.~S., Eplee, R.~E., Jr., Isaacman, R.~B.,
Meyer, S.~S., Shafer, R.~A., Weiss, R., Wright, E.~L., Bennett, C.~L,
Boggess, N.~W., Dwek, E., Gulkis, S., Hauser, M.~G., Janssen, M.,
Kelsall, T., Lubin, P.~M., Moseley, S.~H.~Jr., Murdock, T.~L.,
Silverberg, R.~F., Smoot, G.~F. \& Wilkinson, D.~T. 1990! ApJ!
354! L37-L41;

\refitem
Mather, J. C., Cheng, E. S., Cottingham, D. A., Eplee, R. E., Jr.,
Fixsen, D. J., Hewagama, T., Isaacman, R. B., Jensen, K. A.,
Meyer, S. S., Noerdlinger, P. D., Read, S. M., Rosen, L. P.,
Shafer, R. A., Wright, E. L., Bennett, C. L., Boggess, N. W.,
Hauser, M. G., Kelsall, T., Moseley, S. H., Jr., Silverberg, R. F.,
Smoot, G. F., Weiss, R. \& Wilkinson, D. T.
1994! ApJ! 420! 439-444;
%.!"Measurement of the CMB Spectrum by the COBE FIRAS"

\bookref
Wright, E. L. 1993, in ``Back to the Galaxy'', AIP Conference Proceedings 278,
eds. S. S. Holt \& F. Trimble, p. 193.
%.!"Cold Dust in the ISM?"

\refitem
Wright, E. L. etal. 1992! ApJL! 396! L13;
%.!"Interpretation of the Cosmic Microwave Background Radiation Anisotropy
%.! Detected by the COBE DMR"

\refitem
Wright, E. L., Mather, J.~C., Bennett, C.~L., Cheng, E.~S., Shafer, R.~A.,
Fixsen, D.~J., Eplee, R.~E.~Jr., Isaacman, R.~B., Read, S.~M., Boggess, N.~W.,
Gulkis, S.~G.,  Hauser, M.~G.,  Janssen, M., Kelsall, T.,  Lubin, P.~M.,
Meyer, S.~S., Moseley, S.~H.~Jr., Murdock, T.~L., Silverberg, R.~F.,
Smoot, G.~F., Weiss, R.,  and Wilkinson, D.~T., 1991! ApJ! 381! 200-209;
%.!"Preliminary Spectral Observations of the Galaxy with a 7\deg\ Beam
%.! by the Cosmic Background Explorer (COBE)"

\refitem
Wright, E. L.,
Mather, J. C., Fixsen, D. J., Kogut, A., Shafer, R. A., Bennett, C. L.,
Boggess, N. W., Cheng, E. S., Silverberg, R. F., Smoot, G. F. \& Weiss, R.
1994a! ApJ! 420! 450;
%.!"Interpretation of the COBE FIRAS Spectrum"

\end{document}